# Emergence of heterogeneity in an agent-based model

Wan Ahmad Tajuddin Wan Abdullah

*Deparment of Physics, Universiti Malaya, 50603 Kuala Lumpur, Malaysia*

We study an interacting agent model of a game-theoretical economy. The agents play a minority-subsequently-majority game and they learn, using backpropagation networks, to obtain higher payoffs. We study the relevance of heterogeneity to performance, and how heterogeneity emerges.

## 1. Introduction

In this paper, we are concerned with the notion of heterogeneity in models with networked/interacting elements; we are interested to explore whether heterogeneity of elements is necessary or at all useful to the optimization carried out collectively by the system dynamics. The existence of ecosystems with heterogenous species cooperating towards the collective survivalibity suggests that heterogeneity is perhaps at least advantageous. In econophysics, heterogeneous agent models (e.g. ref. [1]) have been proposed to account for behaviour of markets, and in other models, the emergence of heterogeneity have been observed [2,3].

We would like to investigate the emergence of heterogeneity with the ultimate hope of understanding the underlying physics. We choose the specific context of a model with initially symmetric agents interacting to learn strategies to maximize rewards in an iterated global game without known analytic solutions as yet [4]. The agents learn through individual backpropagation-type neural networks, and we decode learnt strategies from the resulting pattern of synaptic strengths.

## 2. The Model

We chose to work in the context of a model where agents play an iterative game with global interactions, in the sense that payoffs do not depend locally on individual agents' moves separately, but collectively on the moves seen together, and with delayed payoffs, in the sense that they depend on the values of the players' future moves. In particular, subscribing to some economic intuition, we design payoffs in such a way that we reward (near-) minority choices which later on become popular. (This is an extension of the minority game [5]). The unpredictability introduced by having delayed payoffs is hoped to prevent agents all learning calculable optimum behaviour as in models with known solutions, thus to offer a greater chance of producing 'spontaneous' heterogeneity.

$N$ players or agents $k$ each choses one of $n$ options $i$. If $i_k(t)$ is $k$'s choice at timestep $t$, then payoffs $p_i$ for moves $i$ at $t$ is taken to be

$$p_i(t) = \sum_k \delta(i, i_k(t+1)) - \sum_k \delta(i, i_k(t)) \qquad (1)$$

where the Kronecker delta $\delta(x,y) = 1$ when $x = y$ and 0 otherwise.

We choose a system with learning agents rather than an evolutionary system. Agents learn online using a neural network without hidden units, by employing a variant of the back-propagation algorithm [6], or equivalently, Hebbian learning with nonlinearity. For an input $u$ contributing to the output $i_k(t)$, the synaptic strength $T^k_{ui}$ between them changes by

$$\Delta T^k_{ui} = \eta (p_i - h^k_i) f'(h^k_i) + \chi \iota \qquad (2)$$

where $\eta$ is the learning rate, $h^k_i$ is the field for the neuron representing $i_k$,

$$h^k_i = \sum_u T^k_{ui} u, \qquad (3)$$

$f'(x)$ is the differential of the neuron transfer function (the nonlinearity), taken to be

$$f'(x) = 1/(x^2+1), \qquad (4)$$

and $\chi$ is a 'creativity' factor incorporating 'noise' or some 'non-zero temperature' effects via the random number $\iota$ picked from a uniform random distribution between 0 and 1. Initial synaptic values are small random numbers equally likely to be positive or negative. Without learning ($\eta = 0$), option choices are random.

Agents decide on options through a competitive network [7] involving the final layer neurons; in particular, the agent $k$ choses the option $I$ if $h^k_I = \max(\{h^k_i\}_i)$. We further defined the memory $M$ of agents as the number of timesteps back that they remember of payoffs and moves. Cases where $M = 0$ then are equivalent to random choices.

We can vary the amount of information available from which the agents can learn. Inputs $u$ to the field for decision and to the learning may consist only of each respective agent's own previous moves ('self learning'). A more plausible scenario is where $u$ also include previous resulting payoffs ('learning with payoff'), and with information, there is also the case where in addition to these, the previous moves of other agents contribute to the decision-making and the learning ('learning with information'). A model of culture is also explored where there is a direct contribution to the 'knowledge' stored in an agent's synapses from those of the others:

$$T^k_{ui} := \varepsilon \left( \sum_{k' \neq k} T^{k'}_{ui} \right) + (1-\varepsilon) \, T^k_{ui} \qquad (5)$$

with $\varepsilon$ being the 'culture factor' and this inclusion of culture is carried out after learning (with information) is done ('learning with culture').

We have looked for the emergence of heterogeneity in these cases, taking heterogeneity in performances to indicate heterogeneity in gameplay [4]. This paper extends this by searching for heterogeneity in gameplay by looking at learnt synaptic patterns. The long-time behaviour of agents are coded in the respective eventual patterns of synaptic strengths; we search for any systematics in synaptic patterns in cases where heterogeneity in performances emerges.

## 3. Simulation details

We carried out studies on the model using computer simulations. Specifically, 10 agents choose between 5 options. The following parameters were used: initial randomness (maximum magnitude of initial random value for synapses) 0.1, creativity 0.1, culture factor 0.2 and learning rate 0.1. We studied (100 random trials) long-time behaviour (after 1000 timesteps) of cases of no learning, self learning, learning with payoff, learning with information and learning with culture, each with zero memory, and memories of 1, 3 and 7.

We investigated the mean (between agents) average (over timesteps) payoffs obtained by agents, which measures the average performance of the players, and the standard deviation (between agents) of the average payoffs, which gives a measure of variety of the performances. We chose cases where standard deviations are large, and study synaptic patterns possessed by agents in these cases.

## 4. Results and Discussions

We have reported results of studies on the various learning cases previously [4]. Cases of no learning and all cases of learning with zero memory yield typically a well-defined cluster of means of average payoff at around -0.8 and standard deviations between about 0.02 to 0.08, specifying the performance in a completely random situation. Self learning generally betters the average performances without significantly increasing variances with better average performances when only immediately recent states are remembered. Learning with information shows similar behaviour to learning with payoff, both showing better overall performances through higher means, and emergence of variety through increased standard deviations. Learning with culture smears the means and makes them vary substantially, but the standard deviations remain more or less in the band defined by no learning.

We choose the case of learning with payoff with memory 1 which contains events with largest values of standard deviations (Fig. 1). (Interestingly, a linear trend in the standard deviation-mean scatter, which persists with higher statistics, can be detected. This can be further investigated elsewhere.) Fig. 2 shows individual agent payoffs in a sample of 4 of the events with large standard deviations, showing the spread. Although the hint exists, strongly in one case, there is no clear clustering of performances as might be expected when segregation in strategies happen. Also there is a possibility of a 1-versus-the rest kind of clustering or segregation, with one agent outshining the rest.

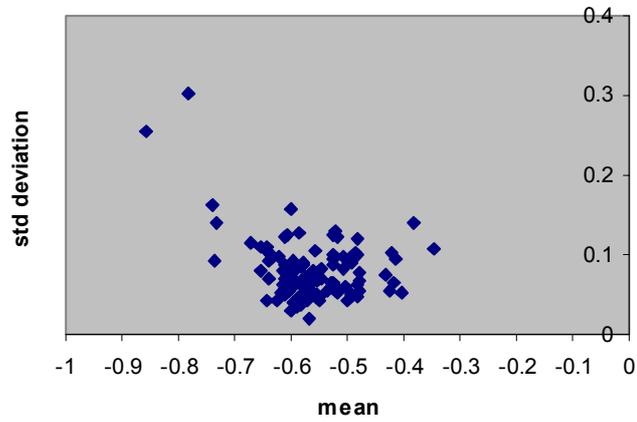

Fig. 1. Standard deviation vs mean of average payoffs for the case of learning with payoff and memory 1.

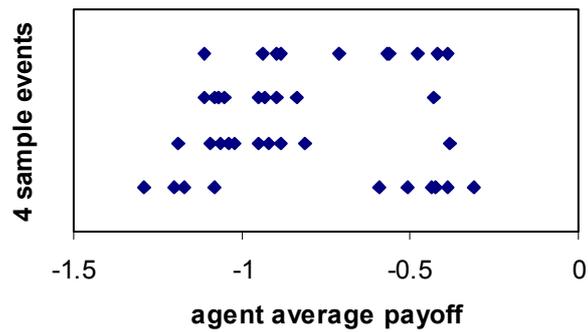

Fig. 2. Individual agent payoffs for events with large variances.

We then select the event with the strong hint of segregation, and compare synaptic patterns of agents within and between the clusters. We did not detect any systematic similarities within clusters nor any systematic differences between clusters. For example, the synaptic patterns for the 2 best performers in the better-performing cluster are shown in Fig. 3.

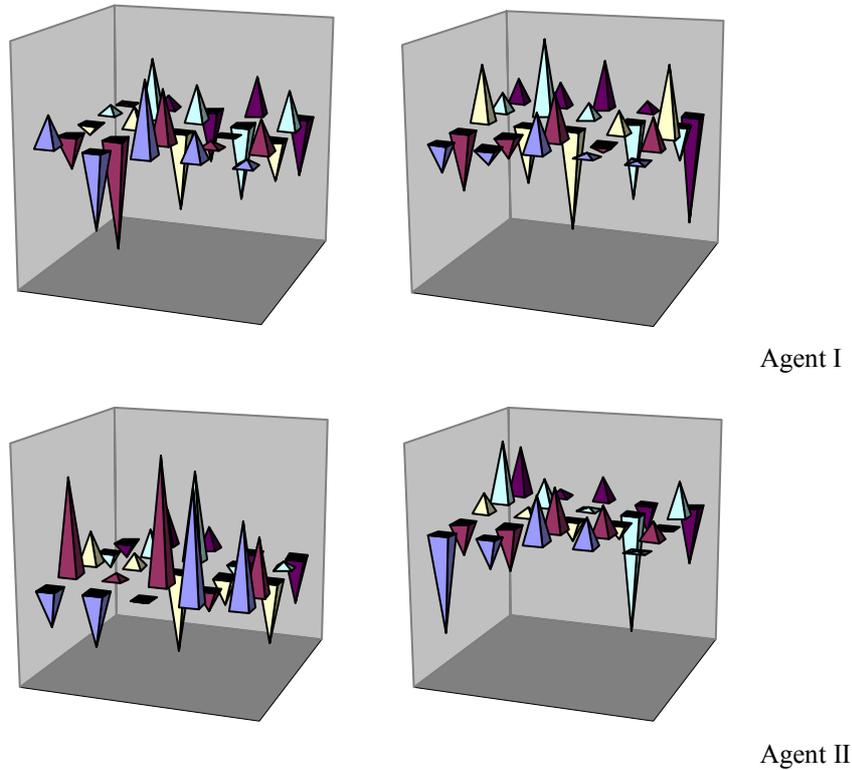

Agent I

Agent II

Fig. 3. Synaptic patterns (synaptic strengths vs option to take vs last option taken, with respect to last payoffs – left, and to last self choices – right) for agents with better average payoffs.

## 5. Conclusions

Through the employment of learning agents in a global game with delayed payoffs, we studied the emergence of heterogeneity. Taking variance in performances to be an indication of heterogenous strategies, we looked at events with largest standard deviations in the average payoffs. Our study did not detect any systematic similarities in strategies (as depicted by end synaptic patterns) between the better performers, nor any systematic differences in strategies between better and poorer performers. However, a future higher-statistics study may possibly uncover such segregating behaviour in the system.